# High-accuracy pointing and sub-second acquisition in a space optical communication terminal with ground-based verification method

Jianmin Wang*, Zhiqian Su, Bin Li, Weiran Zheng, Haochun Gao

**Abstract**—This paper presents and implements a novel space optical communication terminal achieving high-precision open-loop pointing and sub-second acquisition. We further introduce a simple and accurate ground-based performance test method that enables end-to-end verification without the need for in-orbit experiments. To accurately measure open-loop pointing accuracy and acquisition characteristics, we formulate a comprehensive mathematical error model—covering structural misalignments, installation tolerances, and environmental effects—and estimate calibration parameters from stellar observations via a least-squares solution. Field experiments show a >94% improvement in open-loop pointing accuracy, reducing the mean error from 2070.24 μrad to 120.16 μrad, and confirm an average acquisition time of 0.908 s with all trials completed in under 1 s. The method is generalizable to large-range, high-precision optical pointing measurements and astronomical observations.

## 1. INTRODUCTION

With the growing demand for secure, high-speed, large-capacity, and interception-resistant communications, free-space optical communication (FSOC) has emerged as a critical technology for inter-satellite links (ISLs), satellite-to-ground links (SDLs), and deep-space missions. Agencies such as the National Aeronautics and Space Administration (NASA), the European Space Agency (ESA) have conducted multiple in-orbit verification projects to assess FSOC's potential [1]-[3]. In recent years, China has also carried out significant in-orbit experiments, including high-speed inter-satellite laser links and 10 Gbps satellite-to-ground communications [4]-[7]. Furthermore, low-Earth-orbit constellations like Starlink are rapidly deploying ISLs, contributing to the development of a global satellite internet infrastructure [8]-[10].

The core technology of FSOC is pointing, acquisition, and tracking (PAT). Due to satellite platform attitude determination errors, structural and mechanical pointing errors of the optical terminal, orbital uncertainties, and optical calibration errors, the outgoing beam often suffers significant open-loop pointing errors [11]-[14]. This results in a field of uncertainty (FOU) of roughly 5–20 mrad, whereas the signal beam divergence is only 10–50 μrad—a gap of more than two orders of magnitude that makes acquisition inherently challenging [15]-[18]. Consequently, acquisition times for star–star optical links are typically on the order of tens to over a hundred seconds, depending on the scan strategy and initial FOU. Rapid acquisition is operationally essential: in SDL scenarios the contact window with a given optical ground station is typically only a few hundred seconds, so acquisition delays translate directly into lost throughput; in ISLs, periodic obstructions, attitude maneuvers, and dynamic routing require swift re-establishment to sustain end-to-end quality of service (QoS) [13], [19], [20].

At present, the principal strategies for reducing acquisition time are as follows: High-power beacons or beam-expansion optics can cut acquisition to a few seconds when the FOU is narrow, but they raise power/thermal load and eat into link margin—best for short links or platforms with energy headroom; less suitable for long ranges or tight thermal budgets [21]-[23]; Studies [19], [20], [24] shorten acquisition by optimizing scan patterns (e.g., composite-spiral, beacon-less, elliptical-FOU sweeps), but they remain fundamentally limited by the open-loop pointing uncertainty (FOU)—so scan time scales with FOU size—while adding mechanism/calibration complexity and sensitivity to FOU drift; overall latency typically remains higher than dedicated-beacon schemes.

On the other hand, how to verify that a new system possesses open-loop pointing and rapid acquisition capability remains an urgent problem. Current ground verification and test systems primarily use dedicated indoor equipment—exemplified by ESA's System Test Bed (STB) for dynamic metrics and Terminal Test Optical Ground Support Equipment (TT-OGSE) for static/optical checks—which emulate infinity targets with large-aperture collimators. This affords high absolute precision and microradian-class closed-loop PAT evaluation inside a narrow, near-axis uncertainty cone, and supports staged component-to-system integration. However, because the optical path and field of view are set by the collimator, these facilities are intrinsically limited to small-angle (mrad-class) pointing/tracking, cannot deliver degree-scale, wide-area absolute detection or test of open-loop pointing, and cannot directly measure end-to-end acquisition time under realistic outdoor conditions; the hardware is also complex and costly [25] – [28].

Our contributions are twofold: (i) we present a compact terminal architecture that achieves sub-second acquisition [29], [30]; and (ii) we introduce a star-referenced, ground-based scheme that performs degree-scale, wide-FOV open-loop pointing calibration and direct acquisition-time measurement under realistic outdoor conditions. our approach removes the large-collimator constraint, lowers cost/complexity, and delivers statistically robust, μrad-level pointing metrics together with sub-second acquisition, all without in-orbit testing.

## 2. STRUCTURE AND PRINCIPLE

The optical terminal adopts a periscope-type architecture designed to meet the dual requirements of high-accuracy pointing and sub-second acquisition. As shown in Fig. 1, the terminal consists of a two-axis coarse-pointing gimbal (motorized elevation and azimuth drives), a periscope and optical aperture, a stellar-capture channel, a communication channel—comprising relay optics, a coarse-tracking module, a fine-steering module, a transmit module, and a receive module — and the control system. Compared with conventional optical terminals, our design adds a dedicated stellar-capture channel that is spectrally separated from the communication channel by wavelength-selective optics (e.g., a dichroic beam splitter).

The periscope optical path of the optical terminal employs two mirrors positioned at 45° relative to the optical axis, as shown in Fig. 1(b). Each mirror is mounted on a motorized rotation axis equipped with an angle sensor, enabling independent elevation and azimuth rotations over a full 0–360° range. External signal light is reflected by the mirrors and enters the receiving antenna, which is a telescope system with a coaxial four-reflector structure. The star visible light acquisition field of view is 2.5°. This arrangement enables the terminal to redirect its optical axis while maintaining a compact structure, minimizing optical aberrations introduced by steering, and ensuring that the transmit, receive, and stellar-acquisition beams share the same propagation path. An optical channel for stellar acquisition is integrated behind the telescope, where an industrial camera is installed to capture star images.

As discussed above, we adopt a co-boresighted optical design in which the stellar-capture channel and all communication channels share the same optical aperture. Once the stellar-capture channel provides a high-accuracy attitude solution of the optical terminal in the inertial frame, the calibrated line-of-sight offset between the transmit channel and the stellar channel ($\leqslant 1$ μrad) allows the transmit beam's open-loop boresight to be known whenever a star is in view. This removes the error propagation inherent to conventional terminals that rely on star sensors mounted on the spacecraft bus to indirectly infer the terminal's pointing. After calibration, the attitude/azimuth solution error can be better than 2 μrad, yielding a total open-loop pointing error below ~500 μrad (subject to ephemerides, geometry, and link range). By contrast, conventional architectures report pointing errors up to ~4 mrad. With an appropriately chosen transmit beam divergence, sub-second acquisition is therefore achievable.

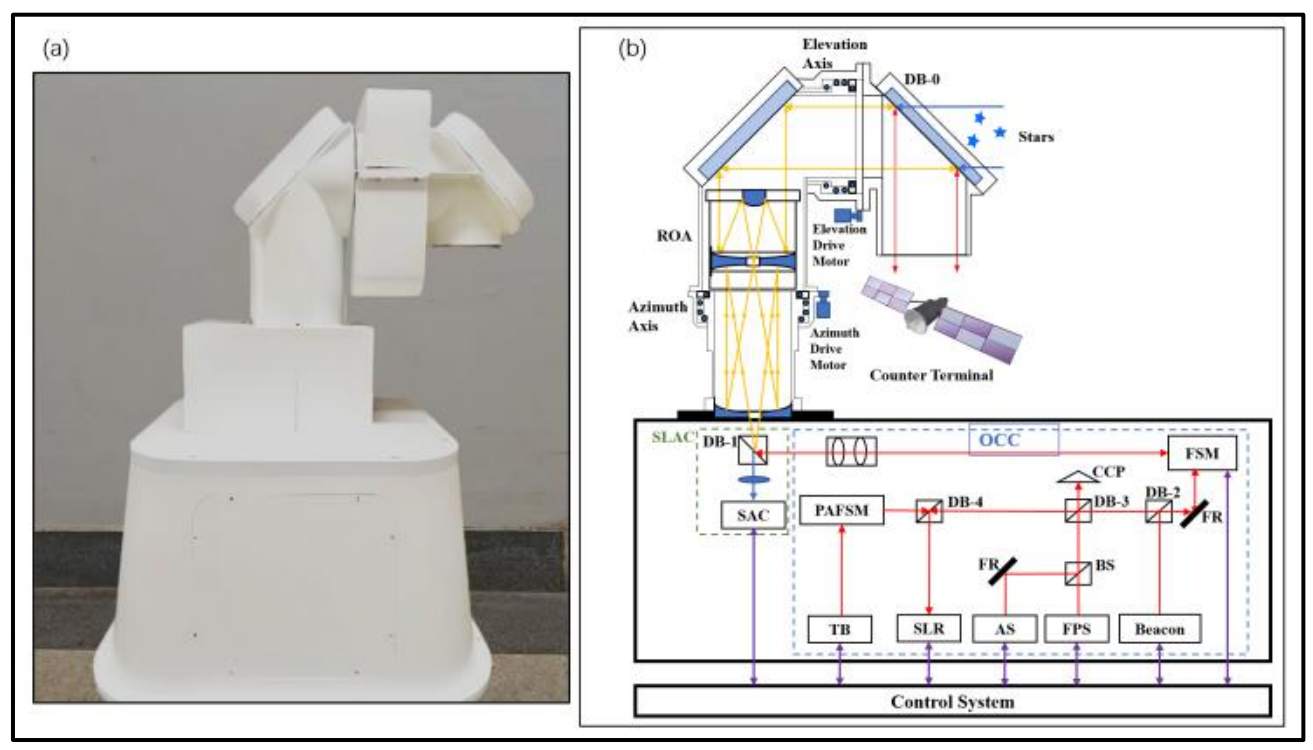

**Fig. 1 - Optical terminal prototype (a) and structural diagram (b). DB-0, 1, 2, 3, 4, Dichroic beam splitters; ROA, reflective optical antenna; SAC, star acquisition camera; SLR, signal light receiver; FPS, fine pointing sensor; CCP, corner cube prism; AS, acquisition sensor; TB, transmitting beam; BS, beam splitter; FSM, fast steering mirror; FR, flat reflector; PAFSM, pointing ahead fast steering mirror; SLAC, star light acquisition channel; OCC, optical communication channel. Shared optical path ensures stellar-based pointing adjustments are directly applied to communication beams, avoiding extra alignment calibration [30].**

Now we explain the principle of the periscope optical path of this terminal and determination. First, this paper defines four mutually related coordinate systems: E-frame — the Earth-centered inertial system with epoch J2000, used to describe the celestial positions of stars; O-frame—the local horizontal frame fixed to the Earth, defined to coincide with the E-frame at the moment of stellar acquisition and thereafter rotating with the Earth. This frame is introduced because the ground experimental platform rotates with the Earth, making it necessary to transform coordinates to a ground-referenced frame; F-frame — the body frame of the optical terminal, whose axes are aligned with the periscope's rotational axes: the $z_F$ and $y_F$ axes are parallel to the terminal's azimuth and elevation axes, respectively, under ideal conditions (the definitions of these axes are detailed later); S-frame—the pixel coordinate system of the star-acquisition camera's imaging plane, which is ideally aligned such that its $x$- and $y$-axes are parallel to those of the F-frame. It should be noted that the star light ray passes sequentially through multiple coordinate frames while representing the same physical ray.

To align with its counterpart terminal, this optical terminal must first determine its own attitude. The attitude-determination procedure can be summarized as follows: the actual direction of a stellar light ray $\boldsymbol{r}_E$ in the E-frame is first transformed—via precise astronomical ephemeris calculations—into the ground-fixed O-frame as $\boldsymbol{r}_O$. It is then mapped to the optical terminal body frame (F-frame) through an attitude determination matrix $\boldsymbol{A}_{ad}$, yielding

$$\boldsymbol{r}_F = \boldsymbol{A}_{ad}\boldsymbol{r}_O. \tag{1}$$

The so-called attitude determination thus amounts to solving for the matrix $\boldsymbol{A}_{ad}$, which defines the terminal's own pointing direction. Subsequently, the dual mirrors in the periscope optical path act effectively as a rotation matrix $\boldsymbol{M}_s$, enabling transformation between the F-frame and the camera's S-frame

$$\boldsymbol{r}_F = \boldsymbol{M}_s\boldsymbol{r}_s. \tag{2}$$

The light ray $\boldsymbol{r}_s$ is then focused onto the imaging plane of the star-acquisition camera and recorded in the camera's pixel coordinate system as $(X_i, Y_i)$, as shown in **Fig. 2**. If the camera principal point is $(X_m, Y_m)$ and the receiving antenna's focal length is $f$, the relationship between the spatial direction of the light ray and the pixel coordinates can be expressed as [31]

$$\boldsymbol{r}_{S,i} = \frac{\begin{bmatrix} -(X_i - X_m)l_p \\ -(Y_i - Y_m)l_p \\ f \end{bmatrix}}{\sqrt{\left(f^2 + \left[(X_i - X_m)l_p\right]^2 + \left[(Y_i - Y_m)l_p\right]^2\right)}}. \tag{3}$$

Here, the subscript $i$ denotes the $i$-th stellar image, and $l_p$ represents the pixel width of star acquisition camera. Thus, we consider $\boldsymbol{r}_S$ as a direction vector in the S-frame, which is related to the pixel coordinates through (3). In the terminal's optical design and calibration, we corrected geometric distortion in the imaging/pointing chain. Measurements and fits show a relative distortion below 0.1% (<1/1000), corresponding to a spot-centroid localization error of <10 μrad. Therefore, distortion can be neglected in the modeling and error budget. As shown in Fig. 2, let φ denote the azimuth angle of the light ray after reflection by the first mirror (Mirror 1)—a rotation about the azimuth axis (the $z_F$-axis of the F-frame) and $\theta$ denote the elevation angle after reflection by the second mirror (Mirror 2)—a rotation about the elevation axis (the $y_F$-axis of the F-frame). Both angles are defined following the right-hand screw rule. In this way, the pointing direction of the light ray in the F-frame can also be described by these two angles. In the F-frame, if the normal vectors of the two mirrors are expressed in terms of $\theta$ and $\varphi$, the normal vector of mirror 1 is $\frac{1}{\sqrt{2}} \cdot [-s\phi \quad c\phi \quad -1]^T$, and the normal vector of mirror 2 is $\frac{1}{\sqrt{2}} \cdot [s\theta \cdot c\phi + s\phi \quad s\theta \cdot s\phi - c\phi \quad c\theta]^T$ . where the superscript $T$ denotes transpose. For brevity, $s$ and $c$ are used to represent the sine and cosine functions, respectively, throughout the paper.

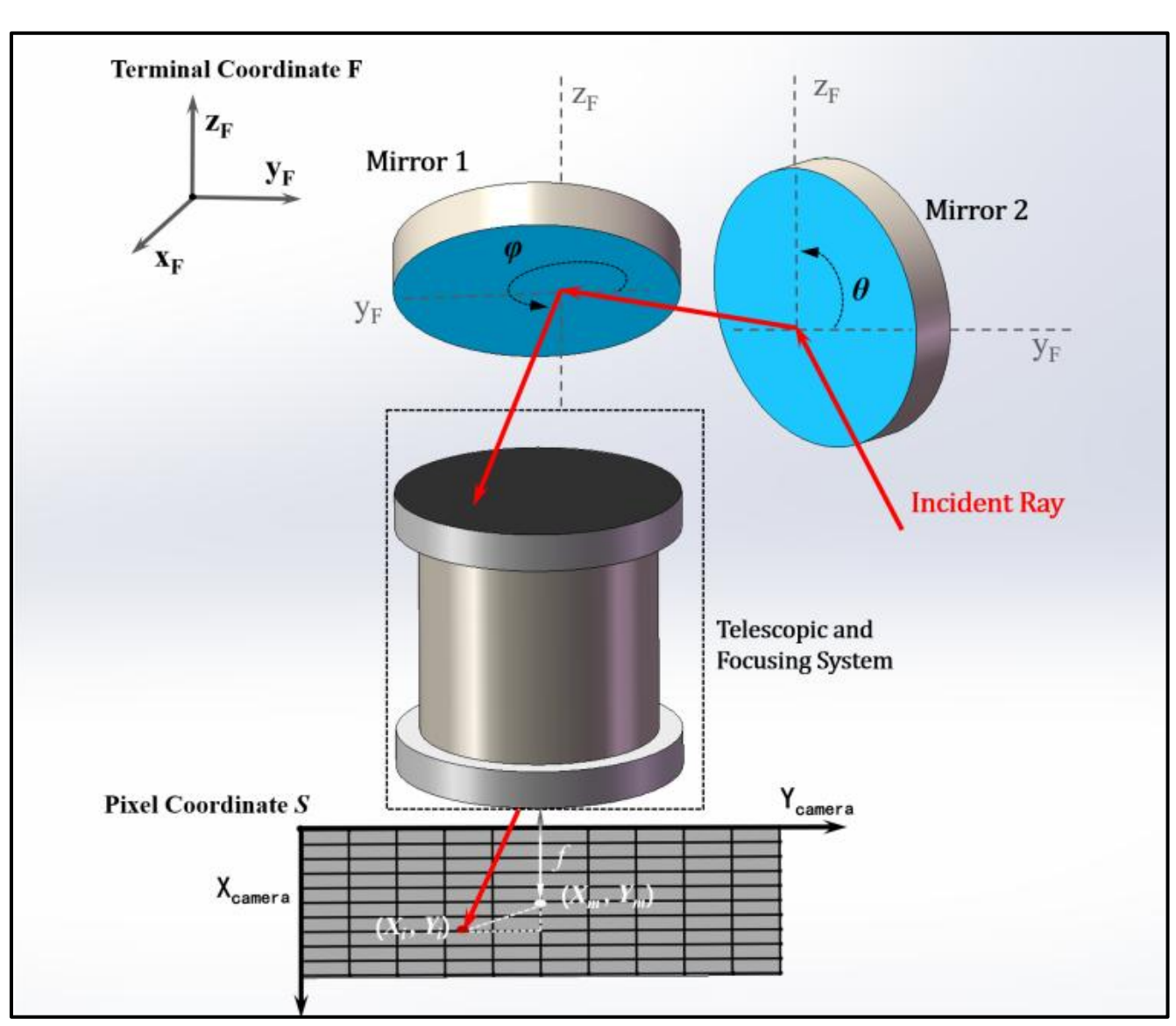

**Fig. 2 - Schematic of the optical path transformation principle of the terminal, indicating the terminal body coordinate system (F-frame) and the camera pixel coordinate system (S-frame). The rotation axes of mirror 1 and mirror 2 are parallel to the $z_F$-axis and $y_F$-axis of the F-frame respectively.**

Using quaternion rotation, the rotation matrix of the periscope optical path, $\boldsymbol{M}_s$ can be derived as

$$\boldsymbol{M}_s = \begin{bmatrix} -s\varphi \cdot s(\theta-\varphi) + c\theta \cdot c\varphi \cdot c(\theta-\varphi), & -s\varphi \cdot c(\theta-\varphi) + c\theta \cdot c\varphi \cdot s(\theta-\varphi), & s\theta \cdot c\varphi \\ c\varphi \cdot s(\theta-\varphi) + c\theta \cdot s\varphi \cdot c(\theta-\varphi), & c\varphi \cdot c(\theta-\varphi) + c\theta \cdot s\varphi \cdot s(\theta-\varphi), & s\theta \cdot s\varphi \\ -s\theta \cdot c(\theta-\varphi), & s\theta \cdot s(\theta-\varphi), & c\theta \end{bmatrix}. \tag{4}$$

Based on the above process, once the star's pixel coordinates are obtained from the camera, a series of coordinate transformations can be applied to compute the true direction of the star, thus determining the open-loop pointing direction of the terminal's outgoing beam. In other words, this completes the conversion of the same light ray through the sequence: $\boldsymbol{r}_S \rightarrow \boldsymbol{r}_F \rightarrow \boldsymbol{r}_O \rightarrow \boldsymbol{r}_E$.

## 3. POINTING ERROR ANALYSIS

Following the procedure described above, the optical terminal's pointing direction can be determined, and the required rotations can be executed. However, in practice, errors inevitably arise. This section analyzes the sources of pointing error for the periscope-type terminal during pointing operations. Other factors that influence the terminal's pointing—such as orbital errors (ephemeris uncertainties) and optical-calibration errors—are irrelevant to our field experiments and are therefore not considered in this study.

The error sources affecting the terminal's open-loop pointing can be divided into two categories: 1. Structural errors, such as mechanical installation errors of the mirrors and rotation axes in the optical path, misalignment of the star-acquisition camera, and uncertainties in key parameters (e.g., the initial readings of the grating angle sensors on the rotation axes within the terminal's body frame, as well as camera-related parameters). 2. Non-structural errors, including those caused by Earth's rotation and atmospheric disturbances.

### A. Structural Errors

#### *Mechanical Errors in the Periscope Optical Path*

We first consider the mechanical errors in the periscope optical path. Because transverse translations of the optical axis do not affect the position of the focused spot under ideal focusing, only angular errors need to be addressed. Ideally, the rotation axis of mirror 1 should be strictly perpendicular to the camera's imaging plane, meaning rotation axis 1 should be parallel to the $z_F$ axis of the terminal's body frame. In addition, the two mirror rotation axes (axis 1 and axis 2) should be perfectly orthogonal. In reality, small deviations occur, referred to as axis orthogonality errors. Similarly, in the ideal installation, both mirrors should be mounted at precisely 45° to the optical axis, with their surface lying in the same plane as their respective rotation axes. In practice, deviations from this orientation arise, known as mirror installation errors. Each mirror has two rotational degrees of freedom, referred to here as the meridional direction and the sagittal direction. The meridional direction is the rotation around an axis perpendicular to both the incident and outgoing beams in the ideal case, while the sagittal direction refers to the orthogonal rotational direction.

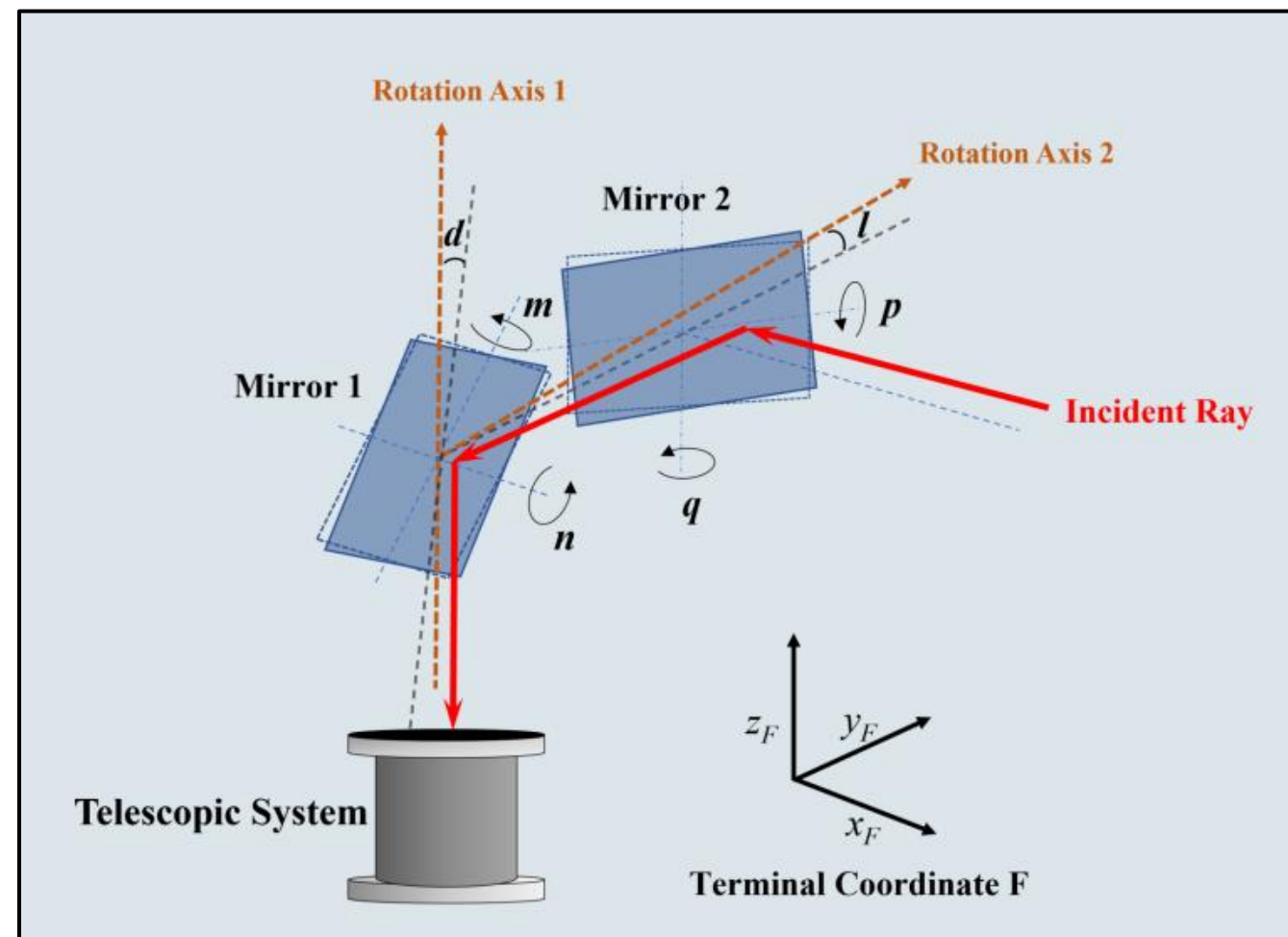

**Fig. 3 - Schematic of mechanical errors in the periscope optical path, showing the planes of the two mirrors, their respective rotation axes, and the various installation error angles of both the axes and the mirrors.**

We now examine how the rotation matrix $\boldsymbol{M}_s$ of the periscope optical path changes when the above mechanical errors are taken into account. First, we assume that the orthogonality error angle between rotation axis 1 and the ideal $z_F$ direction is $d$, the orthogonality error angle between rotation axis 2 and rotation axis 1 is $l$, the installation error angles of mirror 1 in the sagittal and meridional directions are $m$ and $n$ respectively, and the installation error angles of mirror 2 in the sagittal and meridional directions are $p$ and $q$ respectively, as shown in Fig. 3. By substituting the above angles into the original expressions for the mirror normal vectors, we obtain the normal of mirror 1 as

$$\begin{bmatrix} c\varphi \cdot cn \cdot c(\pi/4 + \mathrm{d} + \mathrm{m}) - s\varphi \cdot sn \\ s\varphi \cdot cn \cdot c(\pi/4 + \mathrm{d} + \mathrm{m}) + c\varphi \cdot sn \\ -cn \cdot s(\pi/4 + \mathrm{d} + \mathrm{m}) \end{bmatrix}, \tag{5}$$

and that of mirror 2 as

$$\begin{bmatrix} f_{2x} \\ f_{2y} \\ f_{2z} \end{bmatrix}, \tag{6}$$

where the components are:

$f_{2x} = c\theta \cdot c\varphi \cdot c(d + l) \cdot cq \cdot s(\pi/4 + p) - c\varphi \cdot cq \cdot c(\pi/4 + p) \cdot c(d + l) - s\theta \cdot c\varphi \cdot sq \cdot s(d + l) + s\theta \cdot s\varphi \cdot cq \cdot s(\pi/4 + p) + c\theta \cdot s\varphi \cdot sq$;

$f_{2y} = c\theta \cdot s\varphi \cdot s(d + l) \cdot cq \cdot s(\pi/4 + p) - s\varphi \cdot cq \cdot c(\pi/4 + p) \cdot c(d + l) - s\theta \cdot s\varphi \cdot sq \cdot s(d + l) - c\varphi \cdot c\theta \cdot sq - c\varphi \cdot s\theta \cdot cq \cdot s(\pi/4 + p)$;

$f_{2z} = s(\pi/4 + p) \cdot c(d + l) \cdot c\theta \cdot cq + cq \cdot c(\pi/4 + p) \cdot s(d + l) - s\theta \cdot c(d + l) \cdot sq$. (7)

applying the quaternion method then yields the corrected transformation matrix mapping the light path from the O-frame to the F-frame.

$$\boldsymbol{M}_{s,c} = \begin{bmatrix} R_{11} & R_{12} & R_{13} \\ R_{21} & R_{22} & R_{23} \\ R_{31} & R_{32} & R_{33} \end{bmatrix}. \tag{8}$$

Each $R_{ij}$ is a function of $\theta$, $\varphi$, and the aforementioned angles: $R_{ij} = f_{ij}(d, l, m, n, p, q, \theta, \varphi)$, the full algebraic expansion is omitted here for brevity.

As shown in (4), by applying the corrected transformation matrix $\boldsymbol{M}_{s,c}$ and the ideal transformation matrix $\boldsymbol{M}_s$ to the same direction vector, the corrected direction vector and the ideal direction vector can be obtained. Through mathematical derivation, the relationship between the angular deviation of these two vectors and the variations in $\theta$ and $\varphi$ can be established: 1. when the elevation angle $\theta$ is fixed and the azimuth angle $\varphi$ varies, the magnitude of the angular deviation vector remains constant, and its direction rotates uniformly along the conical surface of the theoretical pointing as $\varphi$ changes; 2. when the azimuth angle $\varphi$ is fixed and the elevation angle $\theta$ varies, the direction of the error vector remains unchanged, while its magnitude follows a trigonometric variation pattern (as illustrated by the simulation in Fig. 4). Based on this analysis, the combined effect of the six mechanical dynamic error parameters can be simplified into a model containing only three coefficients. Therefore, the angular separation between the corrected and ideal directions can be represented by a trigonometric function:

$$J_e = \varepsilon_1 \cdot \sin(2\theta + \varepsilon_2) + \varepsilon_3. \tag{9}$$

Subsequent error calibration can then use stellar data to construct an equation set and solve for these three parameters, thereby compensating for the mechanical installation errors of the periscope optical path.

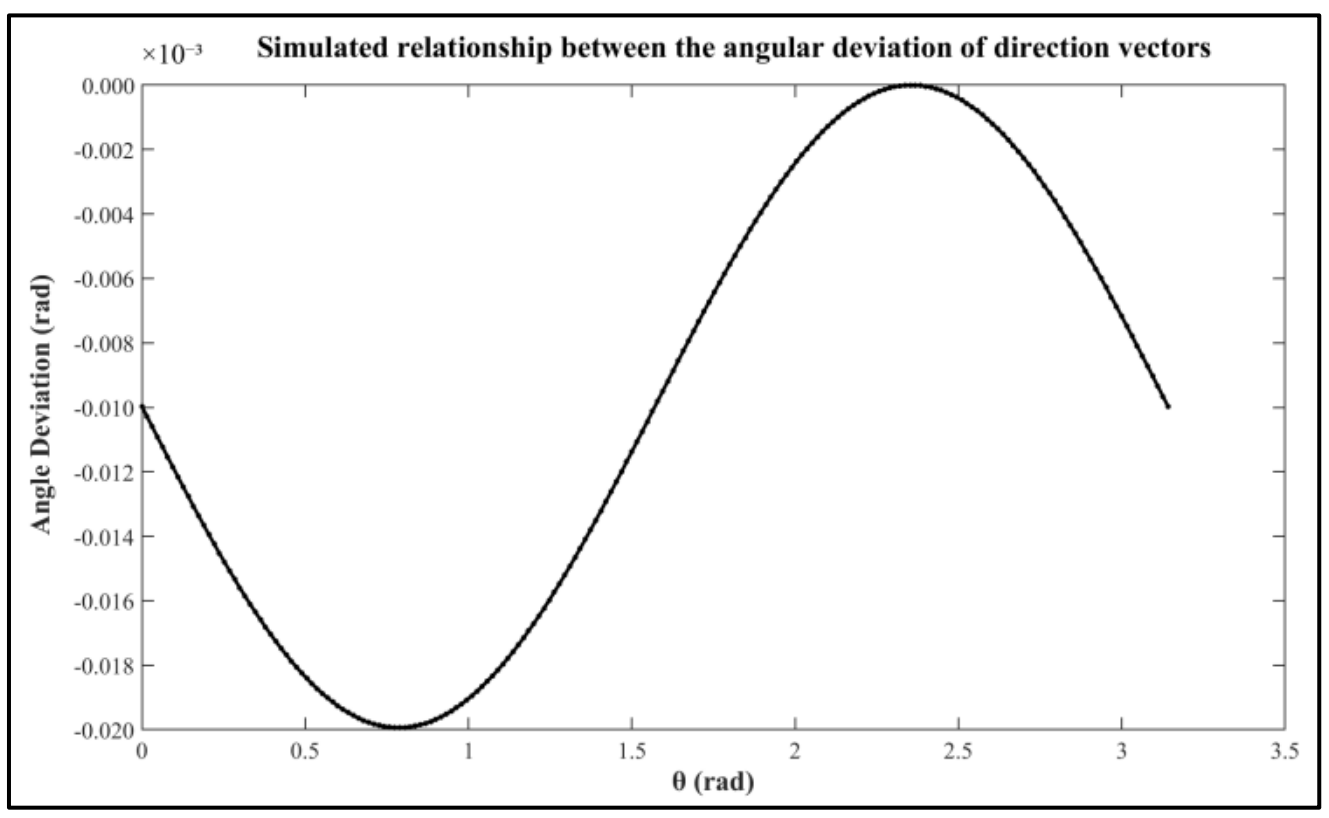

**Fig. 4 - Simulated relationship between the angular deviation of direction vectors obtained by applying the corrected transformation matrix $\boldsymbol{M}_{s,c}$ and the ideal transformation matrix $\boldsymbol{M}_s$ and the variations of $\boldsymbol{\theta}$ and $\boldsymbol{\varphi}$. The coordinate values shown are determined by the parameters set in the simulation and are for illustrative purposes only, not actual measurements.**

***Star-Acquisition Camera Installation Error***

When the star-acquisition camera is correctly installed, its imaging plane should be strictly perpendicular to the optical axis. If the imaging plane remains perpendicular to the optical axis but undergoes translation along the optical axis (axial translation) or translation perpendicular to the optical axis (radial translation), the stellar image will exhibit scaling or overall shift, respectively; however, neither of these translations affects the accuracy of attitude determination. In contrast, if the imaging plane is not perpendicular to the optical axis (i.e., there is an installation tilt), its impact on attitude determination must be quantitatively analyzed.

If the camera plane is tilted relative to the axis-perpendicular direction, the imaging process effectively causes the stellar light ray $\boldsymbol{r}_{S,i}$ to rotate around the optical axis by a certain angle. For small tilt angles, this effect can be represented by a small-angle rotation perturbation matrix

$$\boldsymbol{R}_{cam} \approx \begin{bmatrix} 1 & 0 & \delta\varphi_y \\ 0 & 1 & -\delta\varphi_x \\ -\delta\varphi_y & \delta\varphi_x & 1 \end{bmatrix}. \tag{10}$$

Here, $\delta\varphi_x$ and $\delta\varphi_y$ represent the small angular deviations of the light ray from the ideal direction along the $x$- and $y$-axes, respectively. Thus, the stellar light ray direction vector becomes

$$\boldsymbol{r}'_{S,i} = \boldsymbol{R}_{cam}\boldsymbol{r}_{S,i}. \quad (11)$$

Because the rotational perturbation acts on all stellar light rays simultaneously, and since star map recognition depends on calculating the angular distance between two rays as $\arccos(\boldsymbol{r}_{S,i}^T \cdot \boldsymbol{r}_{S,j})$, when all rays are subject to the same rotational perturbation, the angle between them does not change:

$$\arccos(\boldsymbol{r}'^T_{S,i} \cdot \boldsymbol{r}'_{S,j}) = \arccos\left((\boldsymbol{R}_{cam}\boldsymbol{r}_{S,i})^T \cdot (\boldsymbol{R}_{cam}\boldsymbol{r}_{S,j})\right) = \arccos(\boldsymbol{r}_{S,i}^T \cdot \boldsymbol{r}_{S,j}). \quad (12)$$

Therefore, a tilt of the image plane relative to the optical axis does not change the angles between the rays. However, in practical star-map identification we compute inter-star angles directly from pixel coordinates; such tilt deforms the triangle formed by three stars and thus biases the estimated angular separations. This effect will be analyzed further in the subsequent discussion of pointing error.

***Initial Readings of the Angle Sensors***

The grating angle sensors mounted on the rotation axes output corresponding angle readings as the axes rotate, but their readings must be related to the actual angular state of the axes within the terminal's body coordinate system (F-frame). When the two rotation axes of the periscope optical path are aligned with the $z_F$-axis and $y_F$-axis of the F-frame, the initial readings of the angle sensors are defined as $\theta_0$ and $\varphi_0$. During subsequent experiments, when the actual rotation angles of the axes are $\theta$ and $\varphi$, the real-time readings of the angle sensors can be expressed as

$$\begin{aligned} \theta_r &= \theta_0 + \theta \\ \varphi_r &= \varphi_0 + \varphi. \end{aligned} \quad (13)$$

Therefore, before performing attitude determination, it is necessary to calibrate the initial readings $\theta_0$ and $\varphi_0$. Strictly speaking, these initial values are not error terms, but they are inherent parameters related to the optical path structure and are therefore discussed here.

## B. Non-structural Errors

***Earth Rotation Effect***

During field experiments, the optical terminal is aligned with stellar constellations for attitude determination. During this process, the Earth rotates by a certain angle, which affects the terminal's pointing, so the influence of Earth's rotation must be taken into account. As defined earlier, a ground-fixed reference frame O that rotates with the Earth is used, and at the start of star acquisition ($t$=0), the O-frame is aligned with the inertial frame E. Afterward, the O-frame remains fixed to the Earth's surface and rotates with it, while the E-frame remains stationary.

The Earth's angular rotation velocity adopts the value recommended by the International Earth Rotation and Reference Systems Service (IERS) [32]: $\omega_{\text{earth}}$ =7.2921150 $\times 10^{-5}$ rad/s. If the experiment starts at a ground reference epoch $t$=0 and the Earth rotates for a duration $t$, the rotation of the O-frame relative to the E-frame can be expressed in quaternion form as:

$$\boldsymbol{q}_{OE} = \cos\left(\frac{1}{2}\omega_{earth}t\right) + \sin\left(\frac{1}{2}\omega_{earth}t\right) \cdot [\boldsymbol{i},\boldsymbol{j},\boldsymbol{k}] \cdot \boldsymbol{p}_{\boldsymbol{pole}}. \quad (14)$$

Here, $\boldsymbol{p}_{pole}$ represents the direction of the Earth's rotation axis (the pole axis) in the E-frame. By converting this quaternion into the corresponding rotation matrix $\boldsymbol{A}_{OE}$, the J2000 inertial direction $\boldsymbol{r}_E$ from the star catalog can be transformed into the direction in the ground-fixed O-frame at the experiment time

$$\boldsymbol{r}_O = \boldsymbol{A}_{OE}\boldsymbol{r}_E. \quad (15)$$

During the data acquisition period of the experiment, Earth's rotation causes a shift in the apparent direction of stars. If this is not corrected, it will introduce attitude errors on the order of tens of microradians. Therefore, when computing star directions from the catalog, the effect must be calculated based on the actual data acquisition duration and subtracted from the results.

***Atmospheric Refraction Correction***

When starlight passes through the Earth's atmosphere, the atmospheric density gradient causes refraction, producing a slight deviation in the apparent direction of the starlight. This effect is referred to as "atmospheric refraction" or "astronomical refraction." When height-dependent atmospheric models are not considered, the atmosphere can be approximated as a uniform medium, and a simplified model can be expressed as:

$$\theta_{real} = \arccos(n_{atm} \cdot \cos(\theta)). \quad (16)$$

Here, $\theta$ is the observed elevation angle, $\theta_{real}$ is the corrected true incident direction, and $n_{atm}$ is the atmospheric refractive index. Its average value under standard sea-level conditions is generally taken as [33]: $n_{atm} \approx 1.000277$. In this paper, this value is applied for first-order correction of stellar directions at all elevation angles to compensate for the impact of light-path deviation on attitude-matching accuracy.

## 4. THEORETICAL MODEL AND EXPERIMENTAL SCHEME

The experiments comprise two parts—an open-loop pointing accuracy test and an acquisition time test—to respectively quantify the terminal's open-loop pointing performance and verify its sub-second acquisition capability. The entire procedure of experimental scheme is summarized in Fig. 5.

For the hardware parameters: the periscope-type optical telescope adopts a coaxial four-mirror configuration, with a 150 mm entrance aperture and a 2.5° field-of-view (diameter). The peak-to-valley (PV) wavefront error over the full FOV is < 0.1 μm. In the visible-band imaging path to the star-capture camera, the effective focal length is 488.3 mm. The camera uses a STAR1000 sensor with 15 μm square pixels, 1024 × 1024 resolution, and 10-bit grayscale output.

The mechanical rotation axis is driven by an Aerotech® torque motor (model S-130-39-A) with a maximum continuous torque of 2.36 N·m. For angle sensing, we employ a Renishaw® Signum™ optical encoder with a 20 μm grating pitch, 36,000 lines per revolution, and an index (zero) mark. The readhead is model Si-NN-0000 with analog output. The encoder system supports a maximum measurement speed of 1042 rev/min (rpm), well above the speeds used in our experiments. The motor controller, matched to the motor and encoder, is an Aerotech® Ensemble™ ML-MXH.

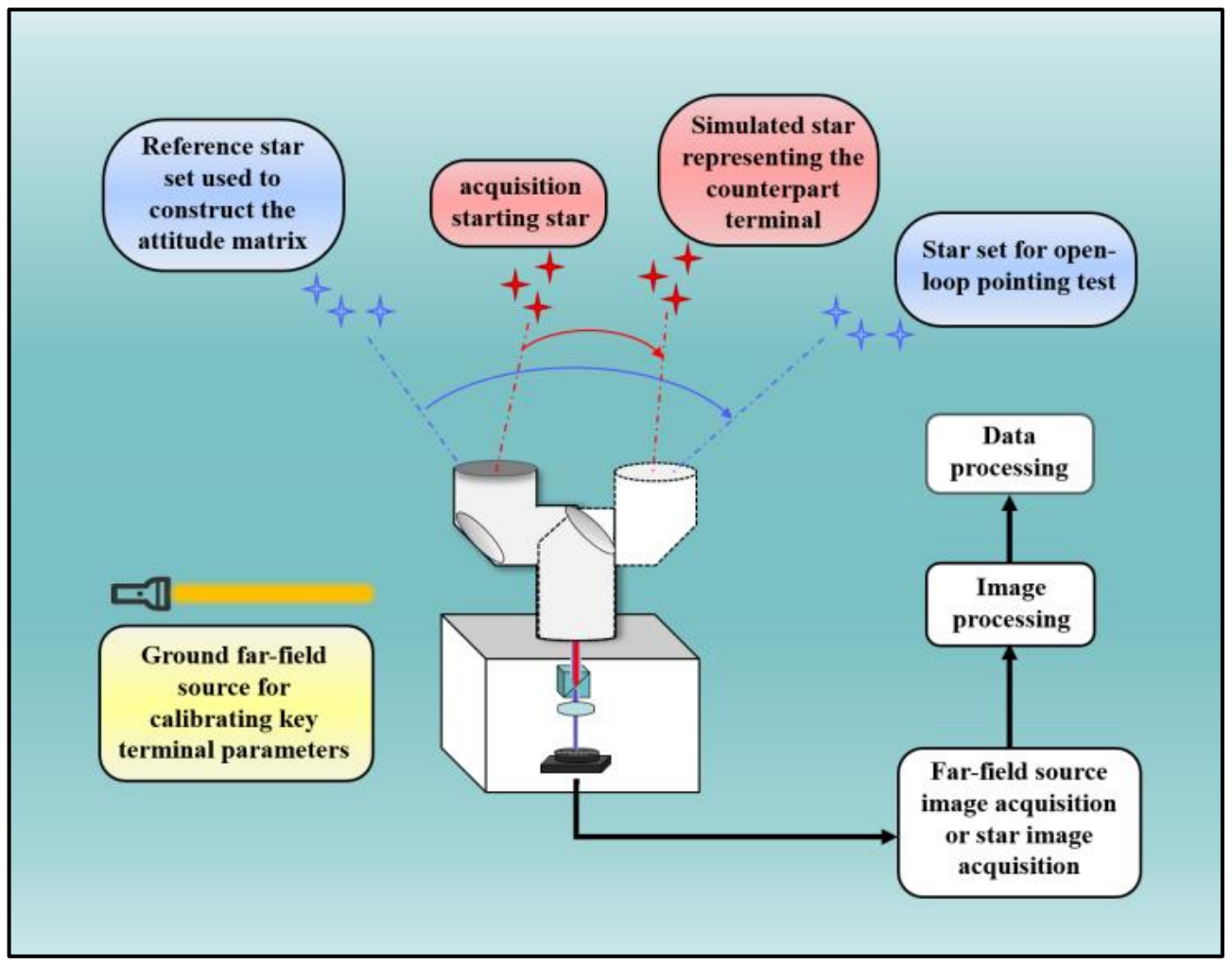

**Fig. 5 - Overall experimental procedure: Open-loop**

The open-loop pointing accuracy test consists of two stages: Stage one—Parameter calibration. A distant ground far-field source is used to calibrate key terminal parameters: star acquisition camera principal point and initial readings of angle sensor. A reference-star set is then observed to build the inertial attitude matrix and establish the mapping from encoder angles to celestial coordinates. Stage two—Optical terminal pointing-error compensation. With the mapping fixed, the periscope/gimbal slews to sequentially image the open-loop pointing star set. The image-processing chain performs centroiding, star identification and matching. The data-processing chain applies first-order corrections for Earth rotation and atmospheric refraction, forms ephemeris residuals, fits and compensates error terms, and finally reports the mean, standard deviation, and RMS pointing accuracy.

For the acquisition-time test, we adopt an equivalent single-terminal "star-to-star switching" method: first acquire one star to initialize the attitude solution, then command the terminal's optical antenna to slew to a second star that emulates the counterpart terminal. From the measured total duration we subtract the pure gimbal slew time ($T_g$), then add the path-propagation time ($T_p$) (computed for a typical two-terminal separation), the FSM convergence time ($T_{FSM}$, measured in the laboratory), and the star-map matching/data-processing time ($T_{mat}$) to obtain the equivalent acquisition time ($T_{eq}$). During testing we log time stamps across the full chain (star-sensor camera sampling ≈ 11 frames per second; FSM operating frequency ≈ 1 kHz with ≈ 16 ms single-step convergence; path delay estimated for 45,000 km), and repeat the "reference-star → target-star" switch across multiple trials to report the distribution and mean of $T_{eq}$.

Now we concentrate on the open-loop pointing accuracy test part. As shown in Fig. 6, the pointing accuracy test consists of two stages: parameter calibration and optical terminal pointing-error compensation. The parameter calibration stage primarily determines the initial readings of the angle sensors and the camera's principal point, serving as preparatory parameters for the next stage. The error measurement–pointing evaluation stage applies a least-squares fitting method to determine error terms and calculate the deviation between theoretical pointing and experimental pointing. Both the experimental scheme and the associated field experiments were carried out in a ground environment. The following sections discuss in detail the mathematical models and data processing for the open-loop pointing test.

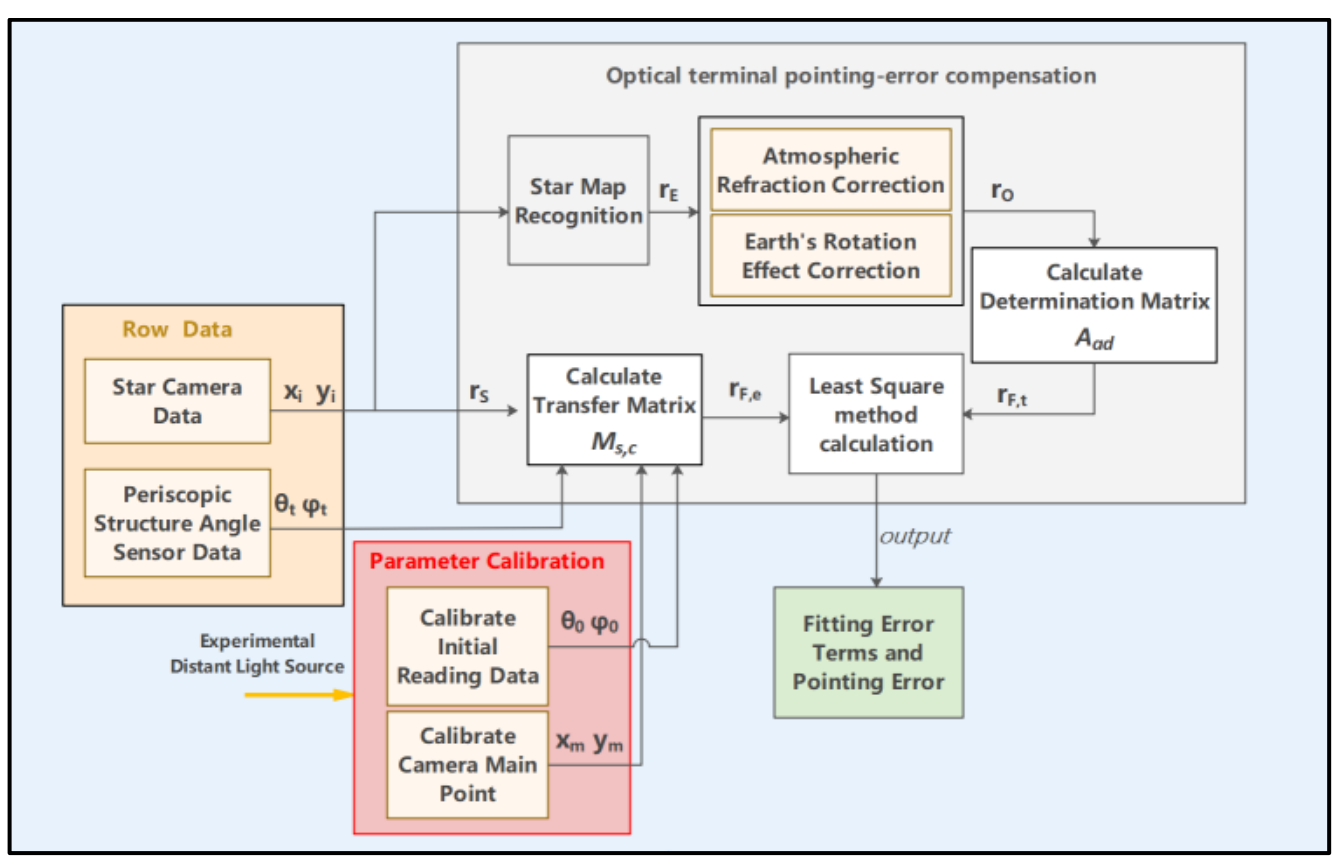

**Fig. 6 - Data processing schematic diagram of open-loop pointing test, the overall scheme divided into (I) parameter calibration and (II) optical terminal pointing-error compensation, ultimately yielding fitted error terms and pointing error data.**

### A. Parameter Calibration

The main purpose of parameter calibration is to determine the principal point of the star-acquisition camera and the initial readings of the angle sensors, providing preparatory parameters for subsequent fine calibration and attitude determination experiments.

Suppose there is a distant, fixed-direction, expanded light source $S_1$, with its incident light rays made as parallel as possible to the optical axis. When the terminal system rotates to a certain actual pointing $(\theta_{1+}, \varphi_{1+})$ (with the angle sensor readings recorded as $(\theta_{r,1+}, \varphi_{r,1+})$), the distant signal ray $\boldsymbol{r}_{F+}$ will focus on a certain point P on the camera image plane. Given that the light source is sufficiently distant, the light reaching the optical terminal is nearly collimated, making point P very close to the camera's principal point. The system is then rotated about the two axes in the opposite directions to the symmetric pointing $(\theta_{1-}, \varphi_{1-})$, where $\theta_{1-} = -\theta_{1+}$ and $\varphi_{1-} = \pi + \varphi_{1+}$. At this time, the angle sensor readings are recorded as $(\theta_{r,1-}, \varphi_{r,1-})$, and the ray $\boldsymbol{r}_{F-}$ should again focus on point P. A ray focusing at the principal point has a direction of $[0,0,1]^T$ in the S-frame. Since point P has a slight positional offset from the principal point $(\varepsilon_{px}, \varepsilon_{py})$, the direction of the ray focusing on point P can be written as $[\varepsilon_{px}, \varepsilon_{py}, 1]^T$. We assume that when the ray focuses on the principal point, the terminal system's pointing is $(\theta_1, \varphi_1)$, and we denote:

$$\begin{aligned}\theta_{1+} = \theta_1 + \Delta\theta_{1+},\ \varphi_{1+} = \varphi_1 + \Delta\varphi_{1+};\\ \theta_{1-} = \theta_1 + \Delta\theta_{1-},\ \varphi_{1-} = \varphi_1 + \Delta\varphi_{1-};\end{aligned} \tag{17}$$

Small quantities $\varepsilon_{px}$, $\varepsilon_{py}$ and $\Delta\theta_{1+}$, $\Delta\varphi_{1+}$, $\Delta\theta_{1-}$, $\Delta\varphi_{1-}$ are substituted into (4), by differentiating $\boldsymbol{M}_s$ with respect to these small quantities and retaining only the first-order terms, the following expressions can be obtained:

$$\begin{aligned}\begin{bmatrix}\Delta\theta_{1+}\\ \Delta\varphi_{1+}\end{bmatrix} &= \begin{bmatrix}-c(\theta_1-\varphi_1) & s(\theta_1-\varphi_1)\\ -s(\theta_1-\varphi_1)/s\theta_1 & -c(\theta_1-\varphi_1)/s\theta_1\end{bmatrix}\begin{bmatrix}\varepsilon_{px}\\ \varepsilon_{py}\end{bmatrix}\\ \begin{bmatrix}\Delta\theta_{1-}\\ \Delta\varphi_{1-}\end{bmatrix} &= \begin{bmatrix}c(\theta_1+\varphi_1) & s(\theta_1+\varphi_1)\\ s(\theta_1+\varphi_1)/s\theta_1 & -c(\theta_1+\varphi_1)/s\theta_1\end{bmatrix}\begin{bmatrix}\varepsilon_{px}\\ \varepsilon_{py}\end{bmatrix}.\end{aligned} \tag{18}$$

We then define

$$\overline{\theta}_1 = \frac{\theta_{r,1+}-\theta_{r,1-}}{2}, \quad \overline{\varphi_1} = \frac{\varphi_{r,1+}+\varphi_{r,1-}-\pi}{2}. \tag{19}$$

Combining (13), (18) and (19), and applying a linear approximation to the sine function, we obtain:

$$-\frac{\theta_{r,1+}-\theta_{r,1-}}{2} = \begin{bmatrix}1 & \sin\overline{\theta}_1\sin\overline{\varphi_1} & -\sin\overline{\theta}_1\cos\overline{\varphi_1}\end{bmatrix}\begin{bmatrix}\theta_0\\ c\varphi_0\varepsilon_{px}+s\varphi_0\varepsilon_{py}\\ -s\varphi_0\varepsilon_{px}+c\varphi_0\varepsilon_{py}\end{bmatrix}. \tag{20}$$

Next, we locate two additional distant light sources $S_2$ and $S_3$ and repeat the procedure, resulting in:

$$\begin{bmatrix}-\frac{\theta_{r,1+}-\theta_{r,1-}}{2}\\ -\frac{\theta_{r,2+}-\theta_{r,2-}}{2}\\ -\frac{\theta_{r,3+}-\theta_{r,3-}}{2}\end{bmatrix} = \begin{bmatrix}1 & \sin\overline{\theta}_1\sin\overline{\varphi_1} & -\sin\overline{\theta}_1\cos\overline{\varphi_1}\\ 1 & \sin\overline{\theta}_2\sin\overline{\varphi_2} & -\sin\overline{\theta}_2\cos\overline{\varphi_2}\\ 1 & \sin\overline{\theta}_3\sin\overline{\varphi_3} & -\sin\overline{\theta}_3\cos\overline{\varphi_3}\end{bmatrix}\begin{bmatrix}\theta_0\\ c\varphi_0\varepsilon_{px}+s\varphi_0\varepsilon_{py}\\ -s\varphi_0\varepsilon_{px}+c\varphi_0\varepsilon_{py}\end{bmatrix}. \tag{21}$$

This forms a 3×3 linear equation system. Solving it yields the initial elevation angle $\theta_0$ and the pixel offset of point P relative to the camera's principal point $(\varepsilon_{px}, \varepsilon_{py})$. Adding this offset to point P's pixel coordinates gives the camera's principal point pixel coordinates.

To determine the initial azimuth reading $\varphi_0$, a near-ground parallel light source $S_g$ is introduced. The azimuth axis of the periscope optical path is locked at $\varphi_{st}$, and only the elevation $\theta$ is varied while several images are taken. Because the azimuth axis is fixed, the light will trace a straight line on the camera plane according to the unit sphere direction-cosine projection model. A least-squares line fitting is performed on all star pixel coordinates $(X_i, Y_i)$

$$Y_i - Y_m = k(X_i - X_m). \tag{22}$$

The slope of the fitted line has the relation:

$$k = -\frac{f}{l_p}\tan(\varphi_0 - \varphi_{st}). \tag{23}$$

From the fitted slope, the initial azimuth reading $\varphi_0$ can be determined.

### B. Optical terminal pointing-error compensation

After completing parameter calibration, high-precision calculations are performed on the captured star images to further solve for system error terms and pointing errors, achieving high-accuracy attitude determination of the periscope terminal in the inertial frame. To enhance stability and representativeness, star data that are evenly distributed, of moderate brightness, and away from the field edges are selected.

The star direction vector $\boldsymbol{r}_O$ in the ground-fixed O-frame is regarded as the “absolute reference”, having been corrected for ephemerides, precession-nutation, atmospheric refraction, and Earth rotation. Two stars approximately 90° apart are selected, and using their directions in O-frame $(\boldsymbol{r}_{O,1}, \boldsymbol{r}_{O,2})$ and their corresponding directions in the terminal F-frame $(\boldsymbol{r}_{F,+}, \boldsymbol{r}_{F,-})$, two sets of orthogonal bases $[\boldsymbol{r}_{O+}, \boldsymbol{r}_{O-}, \boldsymbol{r}_{O,\times}]$ and $[\boldsymbol{r}_{F,+}, \boldsymbol{r}_{F,-}, \boldsymbol{r}_{F,\times}]$ are constructed (the third vector is obtained via the cross product of the first two). The geometric relationship between the two right-handed coordinate systems allows back-solving for an approximate attitude matrix

$$\boldsymbol{A}_{ad} = [\boldsymbol{r}_{F,+}, \boldsymbol{r}_{F,-}, \boldsymbol{r}_{F,\times}][\boldsymbol{r}_{O+}, \boldsymbol{r}_{O-}, \boldsymbol{r}_{O,\times}]^T \qquad (24)$$

Thus, the star catalog direction $\boldsymbol{r}_O$ can be mapped to the terminal body-frame direction (considered theoretical)

$$\boldsymbol{r}_{F,t} = \boldsymbol{A}_{ad}\, \boldsymbol{r}_O. \qquad (25)$$

Given that the calculation of the attitude matrix is based on a finite number of observed stars and the camera plane tilt mentioned earlier, the solution inherently contains small rotation deviations caused by residual attitude disturbances, denoted as$\omega_x$, $\omega_y$, $\omega_z$. The corrected attitude matrix is then

$$\boldsymbol{A'}_{ad} = \begin{bmatrix} 1 & -\omega_z & \omega_y \\ \omega_z & 1 & -\omega_x \\ -\omega_y & \omega_x & 1 \end{bmatrix} \boldsymbol{A}_{ad}. \qquad (26)$$

Accordingly, the corrected attitude matrix gives the terminal’s theoretical pointing as $\boldsymbol{r'}_{F,t}$.

Furthermore, the transformation matrix $\boldsymbol{M}_s$ obtained during parameter calibration maps the star direction $\boldsymbol{r}_S$ in the camera pixel plane into the terminal’s F-frame, yielding the experimentally measured pointing

$$\boldsymbol{r}_{F,e} = \boldsymbol{M}_s\, \boldsymbol{r}_S. \qquad (27)$$

Because the calibration of the camera’s principal point $(X_m, Y_m)$ and the initial angle sensor readings $(\theta_0, \varphi_0)$ has already been completed, $\boldsymbol{r}_{F,e}$ is highly accurate, and its difference from the theoretical pointing $\boldsymbol{r'}_{F,t}$ can be regarded as the pointing error.

Additionally, this pointing error must also consider the camera principal point measurement errors $(\varepsilon_X, \varepsilon_Y)$ and the elevation angle sensor initial reading error $\varepsilon_{\theta_0}$. The azimuth angle initial reading error $\varphi_0$ is ignored because its effect is mainly a slight rotation around the field’s normal, which has negligible influence on the pointing near the center of the field of view. Considering these factors, the angular pointing error between theoretical pointing and experimental pointing is expressed as

$$\phi = \arccos(\boldsymbol{r}'^T_{F,t}, \boldsymbol{r}_{F,e}). \qquad (28)$$

Expanding $\phi$ to the first-order terms of the error parameters $\varepsilon_1$, $\varepsilon_2$, $\varepsilon_3$, $\varepsilon_X$, $\varepsilon_Y$, $\omega_x$, $\omega_y$, $\omega_z$ yields an explicit expression for $\phi$:

$$\phi = \varepsilon_1 \sin[2(\theta + \varepsilon_{\theta_0}) + \varepsilon_2] + \varepsilon_3 - \frac{1}{f}\boldsymbol{r}^T_{F,t} \cdot (\boldsymbol{M}_s \hat{\boldsymbol{e}}_x)\varepsilon_X - \frac{1}{f}\boldsymbol{r}^T_{F,t} \cdot (\boldsymbol{M}_s \hat{\boldsymbol{e}}_y)\varepsilon_Y + \sum_{\{i=x,y,z\}} \left(\frac{\partial \phi}{\partial \omega_i}\right)\Big|_{\omega_i=0} \omega_i\,. \qquad (29)$$

The partial derivative here is expanded in detail as

$$\frac{\partial \phi}{\partial \omega_i}\Big|_{\omega_i=0} = -\frac{(\boldsymbol{r}_{F,e} \times \boldsymbol{r}'_{F,t})_i}{\sin\phi}, \quad (i = x, y, z). \qquad (30)$$

Here, $\hat{\boldsymbol{e}}_x$ and $\hat{\boldsymbol{e}}_y$ are unit direction vectors. Applying a least-squares method to the above equation yields the optimal compensation values for the error terms. Substituting these values back into the expression for $\phi$ produces the final pointing error.

## 5. FIELD EXPERIMENTS

Following the two-stage experimental process, seven field measurement campaigns were conducted to test the terminal’s pointing error and acquisition time. During these tests, different initial attitude states $(\theta, \varphi)$ were covered, including high and low elevation angles and large horizontal rotation angles, ensuring the data fully represent system behavior under diverse initial conditions. The experimental results are summarized in Table 1, which lists the mean pointing error (Mean, μ), the standard deviation of pointing error (STD, 1σ), and the root-mean-square pointing error (RMS, $\sqrt{\mu^2 + \sigma^2}$) before and after calibration and error compensation. Here, the pointing error is calculated using $\phi$ in (29) after substituting the fitted errors, reflecting the residual angular deviation between the terminal’s actual pointing and the theoretical pointing. It should be noted that the fitting process used only part of the observed data to determine the error term values, and these fitted results were then substituted into another independent data set for pointing error evaluation, giving the process stronger statistical significance and methodological reliability. The standard deviation of the pointing error reflects the random pointing fluctuations of the terminal, while the RMS pointing error captures both the overall bias and random variations. Figures 8, 9, and 10 respectively show the results for mean pointing error, standard deviation, and RMS pointing error. The results indicate that, without parameter calibration and error fitting, the terminal suffers from compounded errors resulting in low overall pointing accuracy, with a mean pointing error of 427.15 μrad. After parameter calibration and error fitting, the mean pointing error drops to 120.16 μrad, an improvement of 94.2%, nearly an order of magnitude. Table 2 and Fig. 7 provide additional details for four stellar images captured during the experiment as illustrative examples. Information for other observed stars is available from the authors upon request.

**Table 1 - Experimental results of pointing test before and after calibration**

| Experimental group | Number of Star Point Data | Mean Pointing error (before) [μrad] | Mean pointing error (after) [μrad] | Standard deviation of pointing error (before) [μrad] | Standard deviation of pointing error (after) [μrad] | RMS pointing error (before) [μrad] | RMS pointing error (after) [μrad] |
|---|---|---|---|---|---|---|---|
| 1 | 28 | 3048.92 | 116.17 | 273.05 | 38.34 | 3061.12 | 122.33 |
| 2 | 24 | 3212.33 | 116.42 | 274.74 | 40.72 | 3224.06 | 123.33 |
| 3 | 28 | 2113.38 | 119.29 | 664.62 | 42.85 | 2215.42 | 126.75 |
| 4 | 18 | 1291.61 | 124.84 | 430.73 | 40.02 | 1361.54 | 131.1 |
| 5 | 16 | 1119.46 | 109.64 | 416.03 | 41.33 | 1194.26 | 117.17 |
| 6 | 12 | 1474.85 | 128.73 | 493.47 | 30.4 | 1555.22 | 132.28 |
| 7 | 24 | 2231.09 | 126.05 | 437.41 | 39.09 | 2273.56 | 131.97 |

**Table 2 - Relevant information of Samples of actual star images captured during the experiments**

| No. | Star | Astronomical Information | Observation Information |
|---|---|---|---|
| 1 | Altair<br>(α Aquilae) | Astronomical ID: 97649<br>Absolute Magnitude: 2.2<br>Apparent Magnitude: 0.76<br>Right Ascension: 19h50m46s<br>Declination: 8°52m2s | Azimuth: 289.691464°<br>Zenith Angle: 45.724539°<br>Azimuth Axis Reading: -66.52359<br>Elevation Axis Reading: -76.51725<br>Pixel Coordinates: (551.388, 634.4766) |
| 2 | Mirach<br>(β Andromedae) | Astronomical ID: 677<br>Absolute Magnitude: -0.3<br>Apparent Magnitude: 2.07<br>Right Ascension: 0h8m23s<br>Declination: 29°5m26s | Azimuth: 155.156966°<br>Zenith Angle: 38.922849°<br>Azimuth Axis Reading: 158.62983<br>Elevation Axis Reading: -82.94096<br>Pixel Coordinates: (444.4431, 491.1774) |
| 3 | Aljanah<br>(ε Cygni) | Astronomical ID: 102488<br>Absolute Magnitude: 0.76<br>Apparent Magnitude: 2.48<br>Right Ascension: 20h46m12s<br>Declination: 33°58m10s | Azimuth: 333.98686°<br>Zenith Angle: 39.91607°<br>Azimuth Axis Reading: -21.99739<br>Elevation Axis Reading: -82.26620<br>Pixel Coordinates: (485.8311, 628.5992) |

| No. | Star | Astronomical Information | Observation Information |
|---|---|---|---|
| 4 | Procyon<br>(α Canis Minoris) | Astronomical ID: 37279<br>Absolute Magnitude: 2.68<br>Apparent Magnitude: 0.4<br>Right Ascension: 7h39m18s<br>Declination: 5°13m38s | Azimuth: 164.49477°<br>Zenith Angle: 63.38555°<br>Azimuth Axis Reading: 168.70011<br>Elevation Axis Reading: -58.20388<br>Pixel Coordinates: (620.7439, 556.8584) |

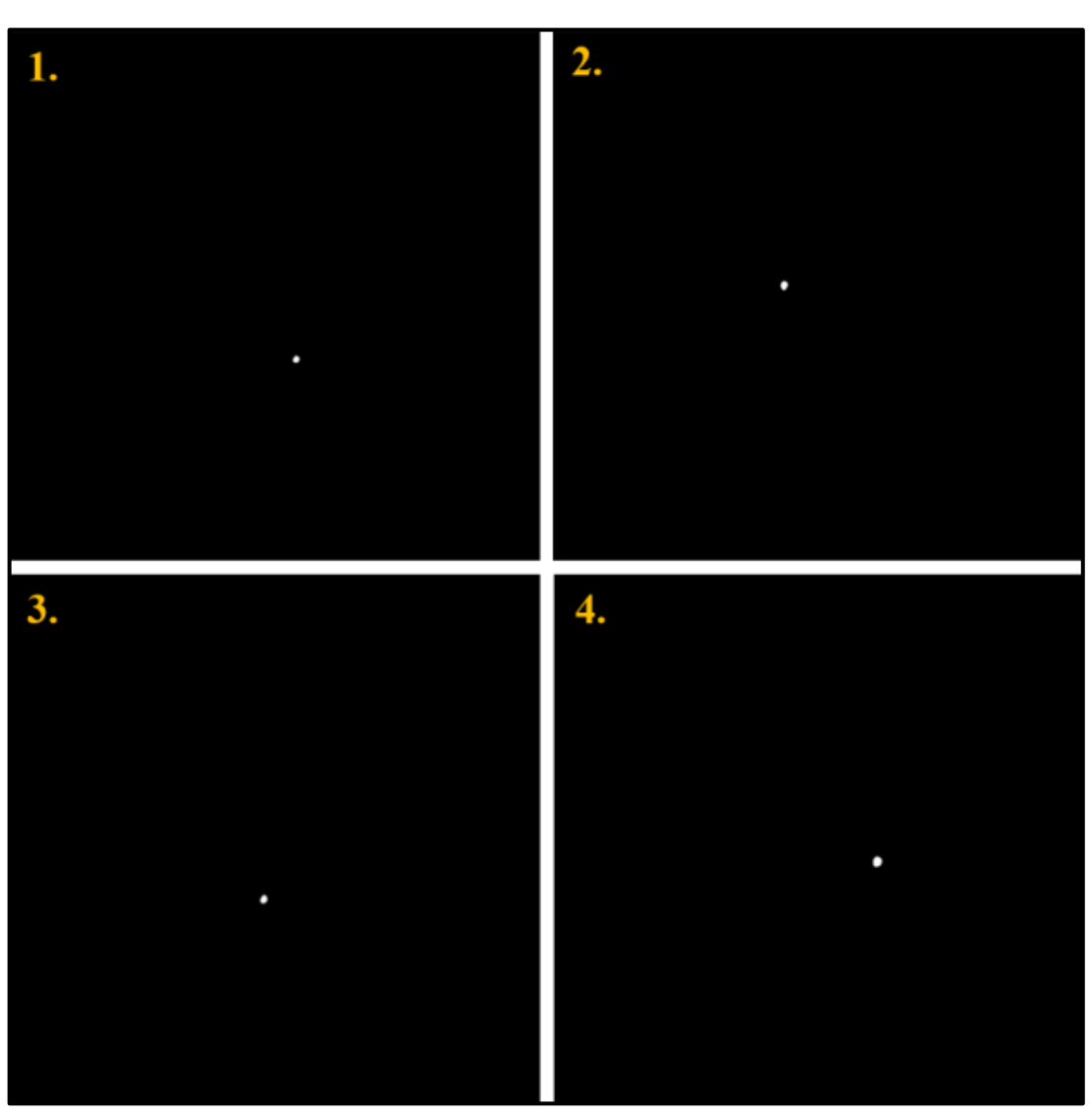

**Fig. 7 - Samples of actual star images captured during the experiments: 1. Altair; 2. Mirach; 3. Aljanah; 4. Procyon.**

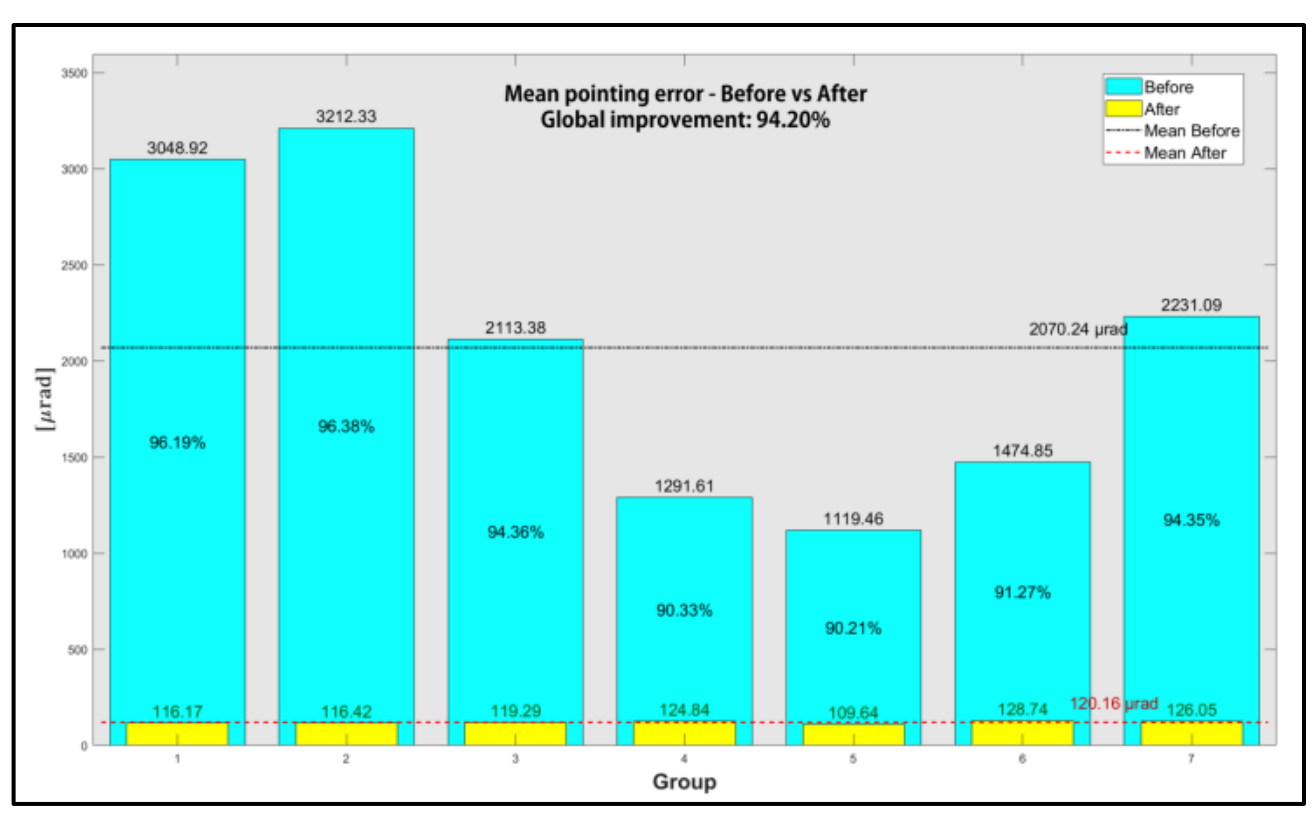

**Fig. 8 - Comparison of mean pointing error before and after calibration.**

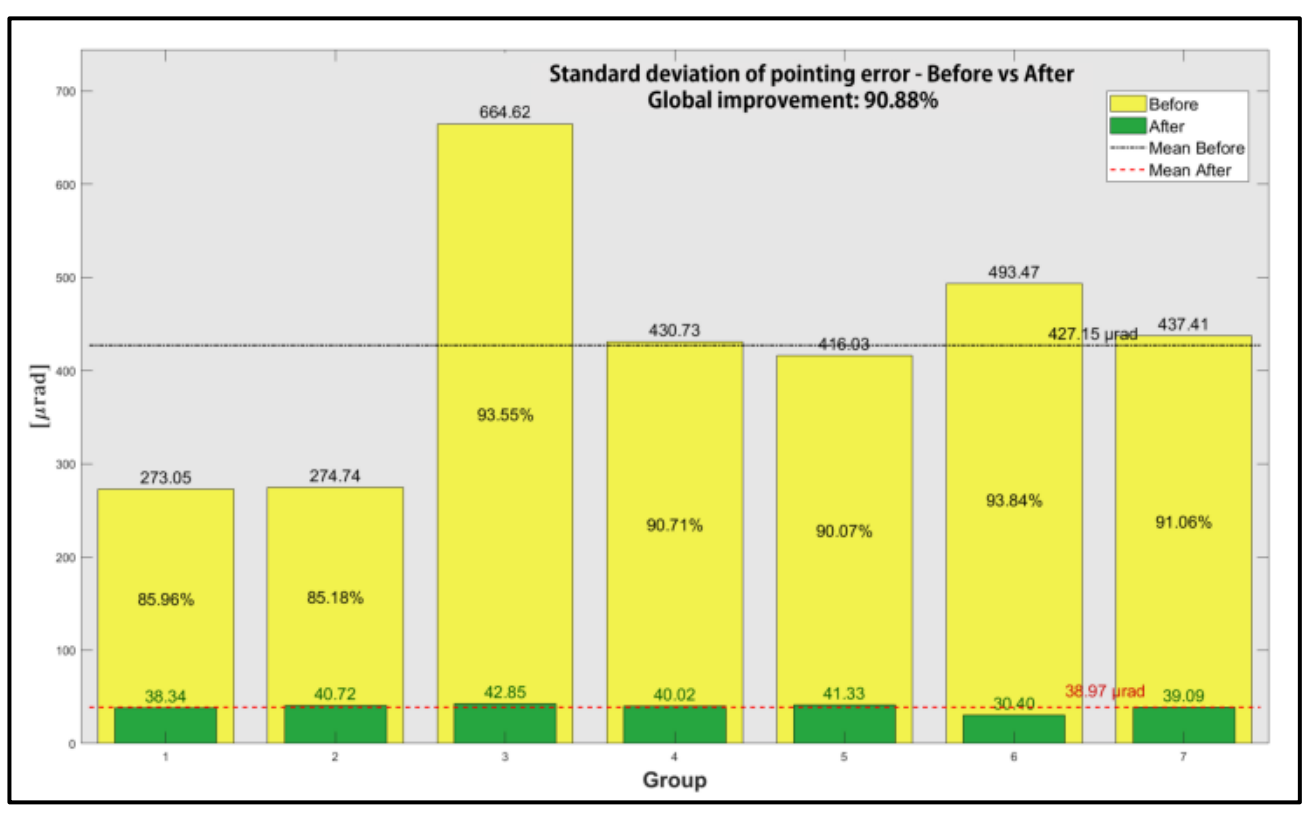

**Fig. 9 - Comparison of pointing error standard deviation before and after calibration.**

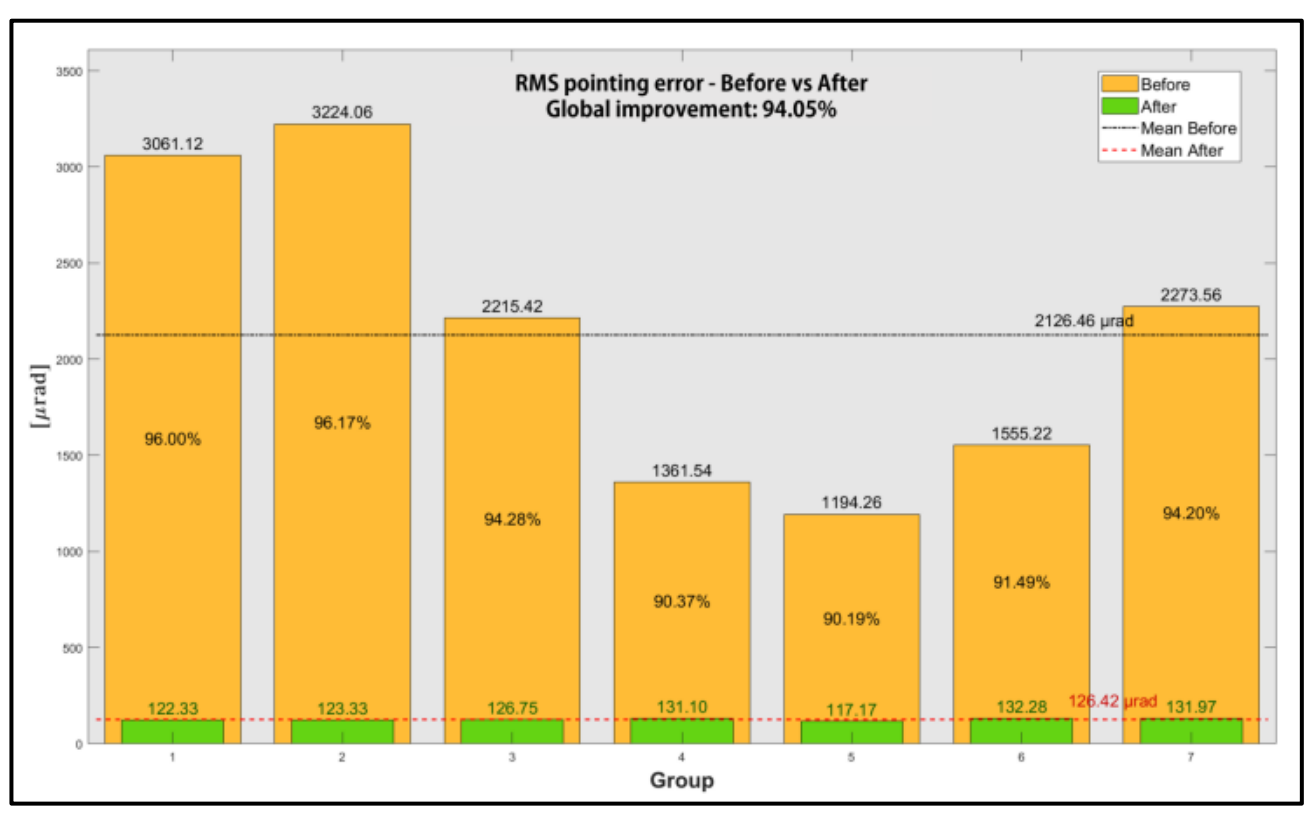

**Fig. 10 - Comparison of RMS pointing error before and after calibration.**

In addition to the histogram-based statistical comparison, Figs. 11-13 provide a more comprehensive view of the pointing error distribution characteristics. Fig. 12 shows the Gaussian fit of the overall pointing error before and after calibration, where the mean error is reduced from 2070.24 μrad to 120.16 μrad and the standard deviation decreases from 427.15 μrad to 38.97 μrad, indicating both a shift toward smaller errors and a significant narrowing of the distribution width. Fig. 11 presents the cumulative distribution function (CDF), in which the 50% quantile decreases from 2363.1 μrad (before) to 118.8 μrad (after), and the 90% quantile drops from 3317.8 μrad to 173.0 μrad, clearly showing that the majority of post-calibration errors are concentrated well below 200 μrad. Fig. 13 shows the global boxplot, including a zoomed-in view for the post-calibration case. The median value decreases from approximately 2350 μrad to 120 μrad, and the interquartile range is reduced by more than an order of magnitude. Together, these figures further confirm the substantial improvement in both pointing accuracy and stability achieved through parameter calibration and error compensation. Throughout the experiments, regardless of the initial rotation axis angles, the model consistently converged and maintained compensation accuracy, demonstrating its stability and general applicability in the operational range.

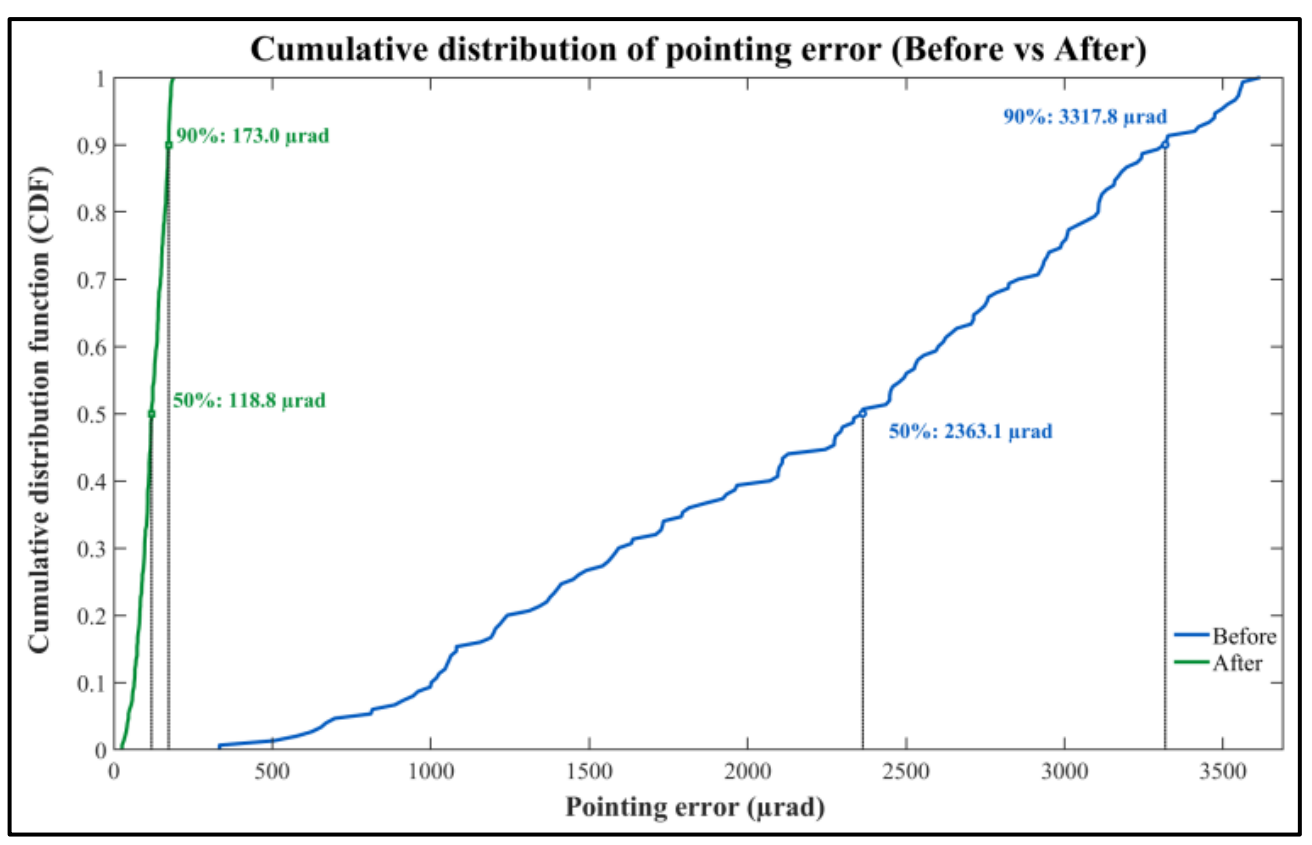

**Fig. 11 - Cumulative distribution of pointing error before and after calibration.**

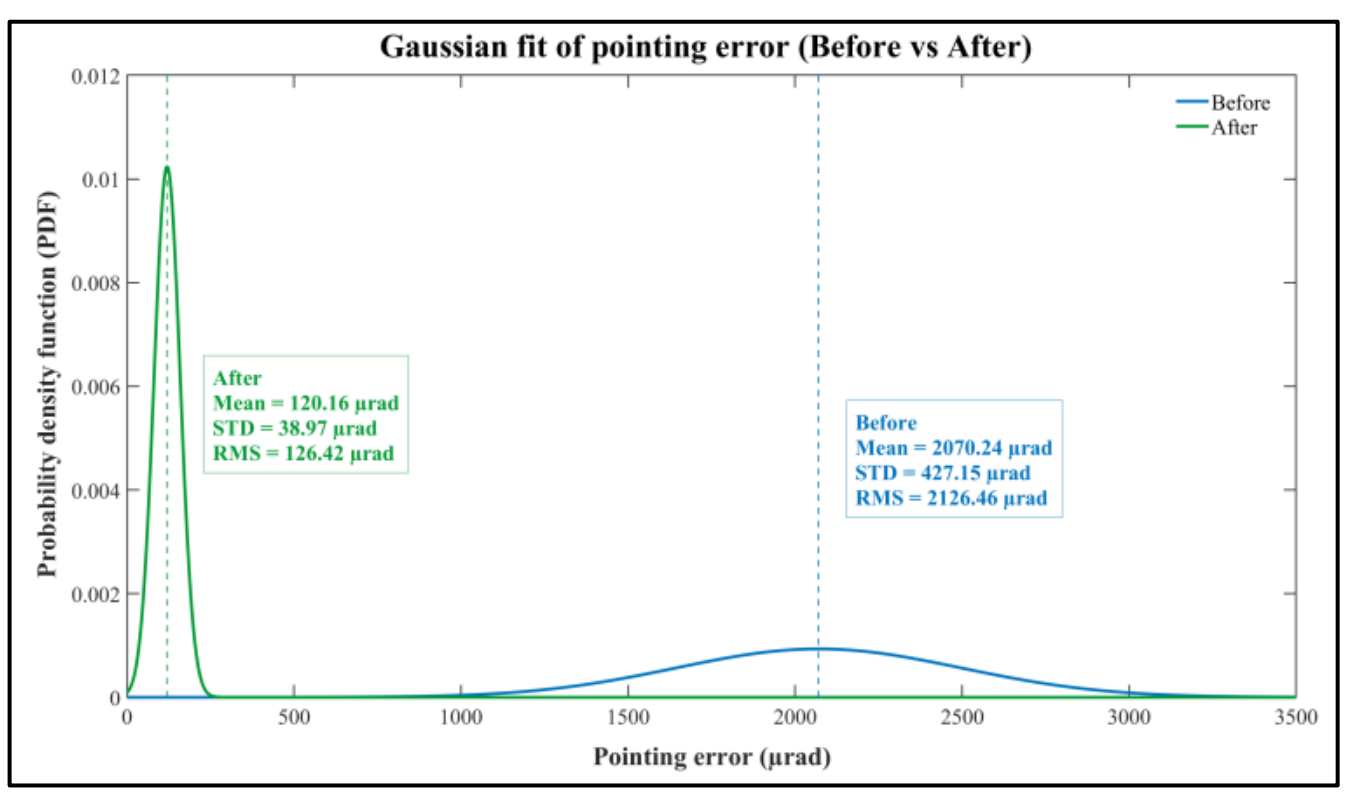

**Fig. 12 - Gaussian fit of pointing error before and after calibration.**

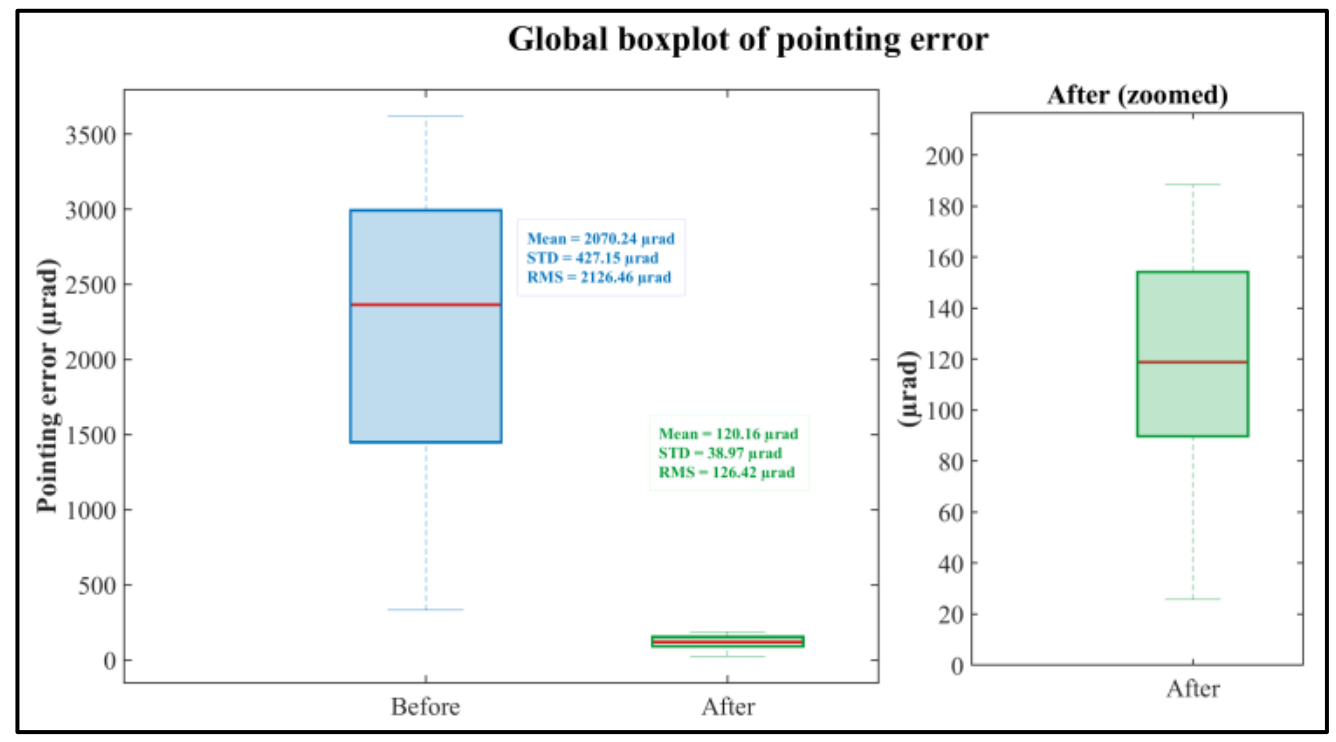

**Fig. 13 - Global boxplot of pointing error before and after calibration.**

In the acquisition time testing experiments, the equivalent acquisition time was measured through single-terminal simulation experiments [25, 26]: multi-star images were first captured to determine the terminal's attitude, the system was then rotated to point at another star to simulate the counterpart terminal, and after subtracting the system's rotation time while adding beam propagation time, fast steering mirror (FSM) adjustment time (This adjustment time was determined through repeated laboratory measurements), and data processing time, the equivalent acquisition time was obtained. Relevant data are provided in Table 3 and Fig. 14. The experiments demonstrate that the optical terminal achieved stable and rapid response in multiple star-switching tasks, with an average equivalent acquisition time of 0.908 s, and all test results remained below 1 s, confirming that the system possesses sub-second acquisition capability and meets the requirements for high-speed link establishment in practical applications.

**Table 3 - Selected stars and measured acquisition times in optical terminal acquisition tests**

| Reference Star | Target Star | Equivalent Acquisition Time [s] | Average Time [s] |
|---|---|---|---|
| Kochab (β Ursae Minoris) | Merak (β Ursae Majoris) | 0.898 | |
| Kochab (β Ursae Minoris) | Dubhe (α Ursae Majoris) | 0.876 | |
| Kochab (β Ursae Minoris) | Arcturus (α Boötis) | 0.903 | 0.908 |
| Kochab (β Ursae Minoris) | Alioth (ε Ursae Majoris) | 0.874 | |
| Kochab (β Ursae Minoris) | Alkaid (η Ursae Majoris) | 0.929 | |

| | | |
|---|---|---|
| Kochab (β Ursae Minoris) | Mizar (ζ Ursae Majoris) | 0.917 |
| Kochab (β Ursae Minoris) | Vega (α Lyrae) | 0.939 |
| Kochab (β Ursae Minoris) | Antares (α Scorpii) | 0.930 |

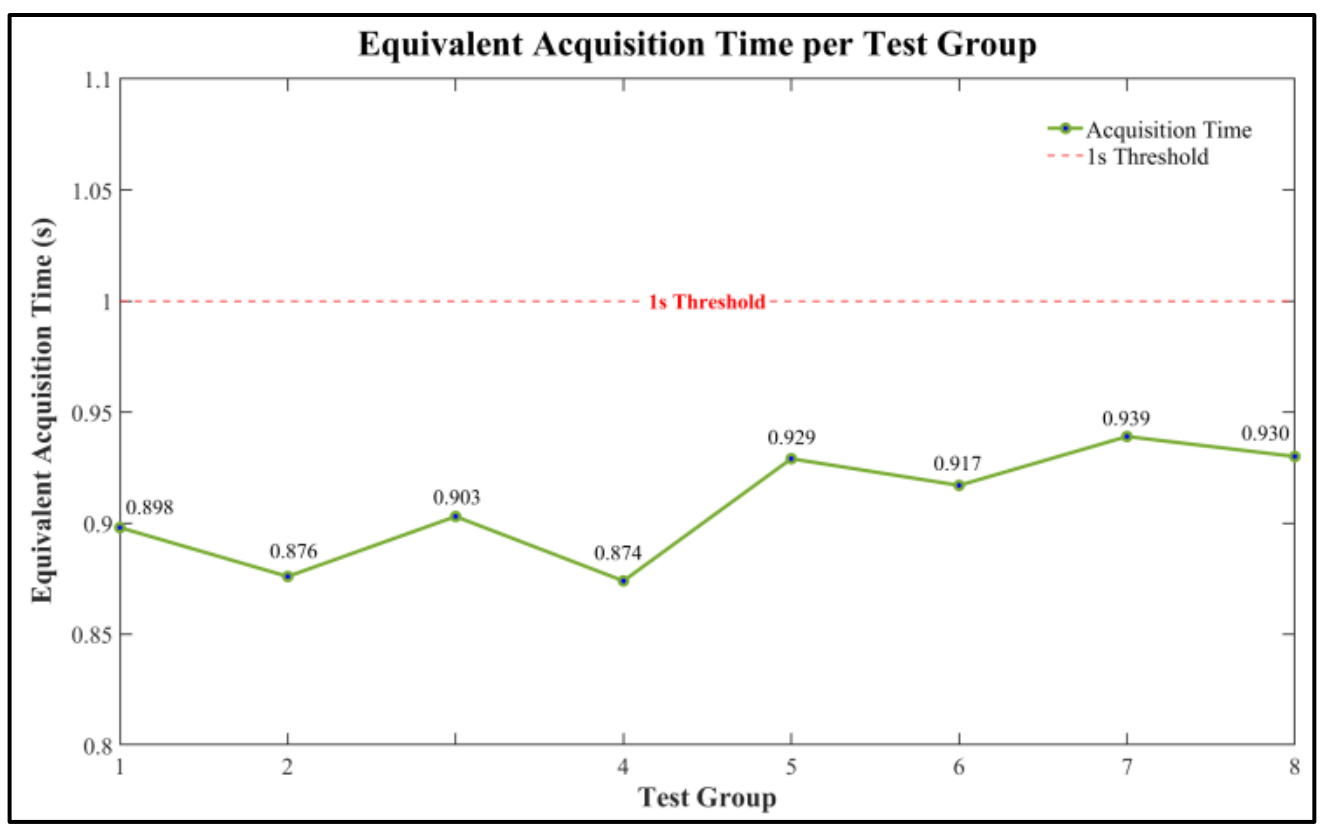

**Fig. 14 - Experimental data of equivalent acquisition time for optical terminal**

## 6. CONCLUSION

To verify the pointing and acquisition performance of the sub-second acquisition periscope-type optical terminal developed by our team, this paper proposes a ground-based field verification scheme. The scheme uses stellar constellations as capture targets, taking their precise positions in the inertial frame as an absolute reference. Combined with optical path structure analysis and simulation calculations, it allows simultaneous evaluation of the terminal's open-loop pointing accuracy and equivalent acquisition time. To address system errors easily introduced by the periscope optical path, a complete mathematical error model was established and corresponding data processing methods were provided, significantly improving open-loop pointing accuracy. Field test results show that after parameter calibration and error estimation, the system's average open-loop pointing error decreased from 2070.24 μrad to 120.16 μrad, an improvement of 94.2%, validating the model's effectiveness and robustness. The acquisition time tests further show that the terminal's average equivalent acquisition time is only 0.908 s, with all tasks completed within 1 s, demonstrating both the feasibility of the proposed field-testing method and the terminal's sub-second fast acquisition capability. The proposed testing scheme also provides valuable reference for ground-based validation of other optical communication systems such as astronomical observation and deep-space optical links.

## 7. APPENDIX

The contents in the appendix and raw data of this paper can be obtained from the authors.

## 8. ACKNOWLEDGMENT

A special thanks to Dr. Shulong Feng from the Changchun Institute of Optics, Fine Mechanics and Physics, Chinese Academy of Sciences, for providing the experimental facilities.